# Evidence of Excitonic Optical Tamm States using Molecular Materials

S. Núñez-Sánchez[1,2], M. López-García[2], M. M. Murshidy[3,4], A. G. Abdel-Hady[5], M. Y. Serry[6], A. M. Adawi[3], J.G. Rarity[2], R. Oulton[2,7], W.L. Barnes[1]

School of Physics and Astronomy, University of Exeter, UK[1]
Photonics Group, Department of Electrical and Electronic Engineering, University of Bristol, UK[2]
Department of Physics and Mathematics, University of Hull, UK[3]
Department of Physics, Faculty of Science, Helwan University, Helwan, Egypt[4]
Yousef Jameel Science and Technology Research Center, The American University in Cairo, Egypt[5]
Department of Mechanical Engineering, The American University in Cairo, Egypt[6]
School of Physics, University of Bristol, UK[7]

## Abstract

We report the first experimental observation of an Excitonic Optical Tamm State supported at the interface between a periodic multilayer dielectric structure and an organic dye-doped polymer layer. The existence of such states is enabled by the metal-like optical properties of the excitonic layer based on aggregated dye molecules. Experimentally determined dispersion curves, together with simulated data, including field profiles, allow us to identify the nature of these new modes. Our results demonstrate the potential of organic excitonic materials as a powerful means to control light at the nanoscale, offering the prospect of a new alternative type of nanophotonics based on molecular materials.

Controlling light beyond the diffraction limit is now a routine matter, and is primarily achieved in the visible and near infrared part of the electromagnetic spectrum by making use of the plasmon modes associated with metallic nanostructures [1]. The key advantages that resonant plasmon modes bring are enhanced electromagnetic field strengths and sub-wavelength field confinement with, for example, a number of strong potential applications in areas such as quantum technology [2]. A number of alternatives to metals have now been explored [3–5], including graphene [6] and other atomically thin materials [7]. Very recently interest has been rekindled in using molecular resonances, especially in the form of molecular excitonic states [8,9]. Such materials may exhibit a strong enough resonant excitonic response that the real part of the associated permittivity becomes negative in the vicinity of the resonance. When this happens the excitonic material may take on a reflective, metallic appearance [10,11]. Early work in this area involved molecular crystals, for which a negative real permittivity was achieved at low temperatures [12]. In the 1970s Philpott and co-workers showed that organic excitonic crystals may support surface exciton-polaritons [12,13], analogous to the surface plasmon-polaritons supported by metals [14]. Recently surface exciton-polaritons supported by dye-doped polymers have been experimentally demonstrated [8, 9], and localized (particle-like) exciton-polaritons theoretically predicted [8,15]. Here we provide the first demonstration that molecular materials can also be used to create another class of nanophotonic mode, specifically Optical Tamm States (OTSs). By extending the range of nanophotonic modes supported by molecular materials, our results open the exciting the prospect of using molecular materials to replace metals as a means to control light at the nanoscale [16]. Such an approach will allow the power of supramolecular chemistry to be harnessed in the design and fabrication of structures to manipulate light in new ways [17], for example using DNA origami [18] .

OTSs are optical modes that occur at the interface between two reflective photonic structures [19], with fields that decay with distance away from the interface that supports them. Demonstration of the existence of these modes has previously been performed using two periodic dielectric stacks [19], or a combination of a periodic dielectric stack and a metallic reflector [20]. Here we show that they can also occur between a periodic dielectric stack and an organic (molecular) reflector, giving a new class of OTS mode, an Excitonic Optical Tamm State (EOTS).

In this Letter we demonstrate the existence of EOTS modes between a truncated 1-D photonic crystal (Distributed Bragg Mirror-DBR) and a polymer layer heavily doped by J-aggregate molecules (the excitonic layer). Light may excite an EOTS when illumination takes place at an incident angle that allows momentum matching to the EOTS to be achieved. The spectrally narrow range over which the metal-like behaviour occurs limits the spectral range over which EOTS may be supported. Below we demonstrate that one can tailor the EOTS dispersion curve and its cut-off wavelengths through the overlap of the momentum-matching mode condition and the wavelength-restricted metal-like properties of the excitonic layer. In what follows, we first describe the characteristic dispersion curves associated with the EOTS based on the optical properties of our excitonic material and on the DBR stopband position. We then present an experimental observation of EOTS excitation in three photonic structures, each showing a different mode cut-off wavelengths.

The propagation condition for an OTS in a multilayer structure DBR/(thin film) requires that the thin film cover layer shows a high reflectivity response within the stop-band of the DBR stack [20]. In the visible metals such as silver can fulfil this condition over the entire visible wavelength range. A dielectric medium doped with excitonic species exhibiting a strong absorption resonance may be modelled using the classical single Lorentz oscillator model,

$$\epsilon_r(\omega) = 1 + \chi(\omega) + \frac{f_0\omega_0^2}{\omega_0^2 + \omega^2 + i\gamma_0\omega}$$

where the resonance frequency ($\omega_0$) corresponds to the exciton transition, $\gamma_0$ is the damping rate, and $f_0$is the reduced oscillator strength; the term $\chi$ takes into account the background susceptibility whilst the reduced oscillator strength is proportional to the concentration of excitonic species. If the oscillator strength is strong enough then the real part of the permittivity will take negative values in a restricted wavelength range, giving a coloured metallic lustre to the organic layer [10,11]. In this work we take the values for the resonance frequency, the damping parameter, and background susceptibility from previous work [8]. For our theoretical study we first considered an excitonic layer with a value of the reduced oscillator strength around unity. Using this oscillator strength, the real part of the permittivity attains negative values over a wavelength band of 100 nm centred on the short wavelength side of the excitonic transition (from 490 nm to just below 590 nm). The excitonic cover layer fulfils the condition of high reflectance only in this restricted wavelength range (see supplementary information, section 1). For the DBR structure we considered a stack formed by 9 pairs of silicon oxide and silicon nitride layers, similar to our experimental samples, but with a central wavelength of 490 nm. The stopband edge of this DBR at normal incidence lies within the high reflectance wavelength band of the excitonic layer.

Figure 1.a shows the simulated P-polarized reflectance (TM) at normal incidence of the DBR stack and two photonic structures supporting OTS, a DBR/(excitonic layer) that supports an EOTS mode and a DBR/(silver layer) that supports a Tamm Plasmon Mode (TP). The DBR shows a high reflectance close to 100 % in the stopband range. However, in the two OTS structures we can observe a sharp dip in the reflectance when the condition of mode excitation is fulfilled. This condition is always observed to occur below the stopband edge of the DBR. If we compare the response of TP structure with the EOTS structure, we observe that in the case of the TP structure the reflectance is close to unity over the whole wavelength range except when the mode is excited. This high reflectance is due to the mirror properties of the silver layer over all of the wavelength range examined. However, in structures supporting EOTSs, reflectance follows the individual DBR response outside the metal-like range of the excitonic material, showing a sharp dip only when the mode condition is achieved. This is because the high reflectance band of the excitonic layer occurs only over a restricted wavelength range, outside this wavelength range the excitonic layer behaves as a simple

dielectric without significantly affecting the DBR response, although the presence of this dielectric layer shifts the Fabry Perot oscillations associated with the DBR to longer wavelengths when compared to the response of the DBR alone.

To confirm the nature of the EOTS mode we need to examine the field intensity profile characteristic of an OTS. Figure 1.b shows the field intensity profile obtained for the TP mode, it is seen to oscillate and to decay with distance away from the interface between the metal and the DBR stack. The decay profile within the DBR is modulated by the periodic variation of the refractive index of the DBR whilst it decays smoothly into the metal medium. Comparing Figure 1.b and Figure 1.c, the electric field profile obtained for the EOTS shows the same features as the TP, confirming that the EOTS has a field confinement that is similar to the TP. In order to confirm that the field profile is only due to the presence of excitonic species embedded on the polymer chain, we have calculated the field profile at the same illumination conditions (wavelength and angle) for a DBR stack covered by an organic thin film layer but with a null oscillator strength. The field intensity profile that we obtain is an oscillatory field that increases with distance away from the interface between the dielectric film and the DBR stack (see Figure 1.d), as is expected when an interference effect is responsible for the high reflectance in the stopband of the DBR stack [21].

Due to the similarities of the TP and EOTS, the dispersion curves of these modes have some properties in common. A TP is always located close to the low-energy stopband-edge [20]. We can use this low-energy stopband-edge to estimate the dispersion curves (see details in supplementary information, section 2). Note that the stopband-edge positions of a DBR stack depend strongly on the polarization and in-plane component of the wavevector. Therefore the mode condition should follow a similar trend to that of the DBR stopband edges. In the case of a TP structure, the mode condition can be met all across the visible wavelength range, however, the EOTS excitation is necessarily limited to the region where the excitonic layer has permittivity values sufficiently negative to obtain a metal-like reflectance. Therefore, the TP shows a continuous dispersion curve while the dispersion curves of the EOTS are truncated.

The truncation (cut-off) wavelengths are delimited by the convolution of the low energy stopband-edge of the DBR and the metal-like reflectance band of the excitonic layer.

Figure 2 shows the dispersion curves obtained for TP and EOTS in structures formed by two different DBR stacks (of central wavelengths 490 nm and 590 nm) with silver and an excitonic (TDBC) reflector. While the TP modes are allowed for all values of the in-plane component of the wavevector in the two different multilayer structures, the EOTS can only be excited over restricted illumination conditions. In the structure with a DBR with a central wavelength around 490 nm we can only excite modes near normal incidence ($k_{||}$=0), whilst for the structure with a DBR central shifted to 590 nm the EOTS mode can only be excited by light impinging close to grazing incidence (large $k_{||}$ values). Therefore, by simply controlling the position of the DBR low-energy stopband-edge relative to the optical properties of the excitonic medium, it is possible to obtain an optical device with a controlled directional response without the need for surface nanostructure.

Three different multilayer structures supporting EOTS were fabricated using different DBR structures but with the same excitonic layer. The DBR mirrors were fabricated using 9 pairs of silicon dioxide and silicon nitride layers deposited by plasma enhanced chemical vapour deposition. The mirrors were designed with central wavelengths ($\lambda_C$) lying in the range 500 nm to 585 nm (DBR-500, DBR-520, DBR-585). The highly reflective excitonic layers were spin cast on top of the DBR mirrors from solutions of the molecular dye TDBC and poly(vinyl alcohol) (PVA) in water. Here we used TDBC molecules as excitonic species (5,6- dichloro-2-[[5,6-dichloro-1-ethyl-3-(4-sulphobutyl)-benzimidazol-2-ylidene]-propenyl]-1-ethyl-3-(sulphobutyl)-benzimidazolium hydroxide, sodium salt, inner salt), with an associated excitonic resonance centred at a wavelength of 590 nm. In appropriate solutions these molecules form J-aggregates; the excitons delocalize in such aggregates leading to (i) a stronger effective dipole moment associated with the exciton, and (ii) a narrower excitonic resonance (arising from exchange narrowing) [22] . This strong dipole moment makes these systems well suited to our purpose since this helps produce an optical response that is strong enough to exhibit a negative permittivity. The TDBC-PVA solutions were prepared by mixing a PVA-water solution with TDBC-water solution by the method explained elsewhere [8]. All TDBC/PVA spun films deposited were designed to yield a 200 nm thickness for the excitonic layer. The optical properties and oscillator parameters of the TDBC/PVA layer were determined in a previous work through reflection and transmission spectroscopy of TDBC/PVA films spun on glass at normal incidence [8].

Angle and polarization resolved reflection maps of the compound structures were obtained using Fourier imaging spectroscopy (FIS) [23]. The samples were illuminated by projecting the output of a fibre coupled to a white light source (Thorlabs OS-L1) through a high numerical aperture objective lens Zeiss 63X ApoFluor (NA=0.75). The reflected pattern generated at the back focal plane of the objective was then scanned with a fibre coupled to a spectrometer (Ocean200+) so as to obtain the angle and polarization resolved reflectance of the structure for all angles within the NA of the objective lens ($\theta^{max}=48^0$ in our case). Angles were then translated into in-plane components of the incident wavevector ($k_{II}$) for comparison with the calculated dispersion curves.

Figure 3 shows the experimental reflectance maps for P-polarized light as a function of energy and in-plane wavevector for three different EOTS structures. Theoretical dispersion curves are plotted on top of the experimental results to visualize the cut-off wavelengths for the different combination of DBRs and excitonic layer. EOTS are supported by all the three structures. For the multilayer structure with a DBR with a central wavelength of 500 nm the EOTS could be only excited for incidence angles below 36 degrees ($k_{||}$~4.7 $um^{-1}$). The dip in the reflectance observed at large wave vectors corresponds to the DBR band edge and not to the presence of an EOTS (see supplementary information). When the central wavelength of the DBR is shifted to lower energies (DBR-520), the dispersion curve is flattened and red shifted while the mode is accessible over a larger in-plane wave vector range (Figure 3.b). In this case it is noticeable that the coupling efficiency is not uniform for all the excitation conditions, in fact the mode is weak at normal incidence while the coupling efficiency is stronger for larger in-plane wave vectors. This effect is due to a fast change of the optical properties of the polymer layer within the wavelength range of the excitonic resonance.

Optical Tamm States are characterized by a splitting between TE and TM polarized modes that increases quadratically as a function of the in-plane wave-vector [19]. Under illumination at the same incident angle (in-plane wave-vector), the wavelength that matches the mode condition will be larger for the TE mode than for TM mode. Therefore, in the case of EOTS, for the same photonic structure we should observe different cut-off wavelengths for TE and TM modes. For comparison purposes Figure 4 shows the S-polarized (TE) reflectance maps for the same three EOTS structures. The TM mode of DBR-500 is restricted to a shorter in-plane wavevector range than the TE mode. Interestingly, the designed structures show a clear splitting between TE and TM modes which in some case might turn into a TE or TM only propagation for a particular wavelength. This was seen for the DBR-585 structure where only a TM mode is observed. These results show that it is possible by design to obtain OTS modes with restrictions on polarization and directionality by selecting the optical properties of the excitonic materials employed.

In conclusion, we reported the first observation of Excitonic Optical Tamm States using an organic layer as a reflector layer, opening a new route to design tailored photonic modes by harnessing the metal-like properties of heavily dye-doped polymer materials to create an all-dielectric nanophotonics. The relatively narrow spectral range of operation might be extended through the use of co-doped thin films containing multiple species [11,24]. The use of excitonic materials in nanophotonics leads to a number of intriguing open questions. A potentially attractive benefit over metals is that the detailed properties of these materials may be controlled by optical (and possibly electrical) pumping, perhaps yielding active functionality. There is also the open question of what happens to quantum emitters embedded in such materials and whether they might be harnessed for quantum technologies [2].

Figure 1

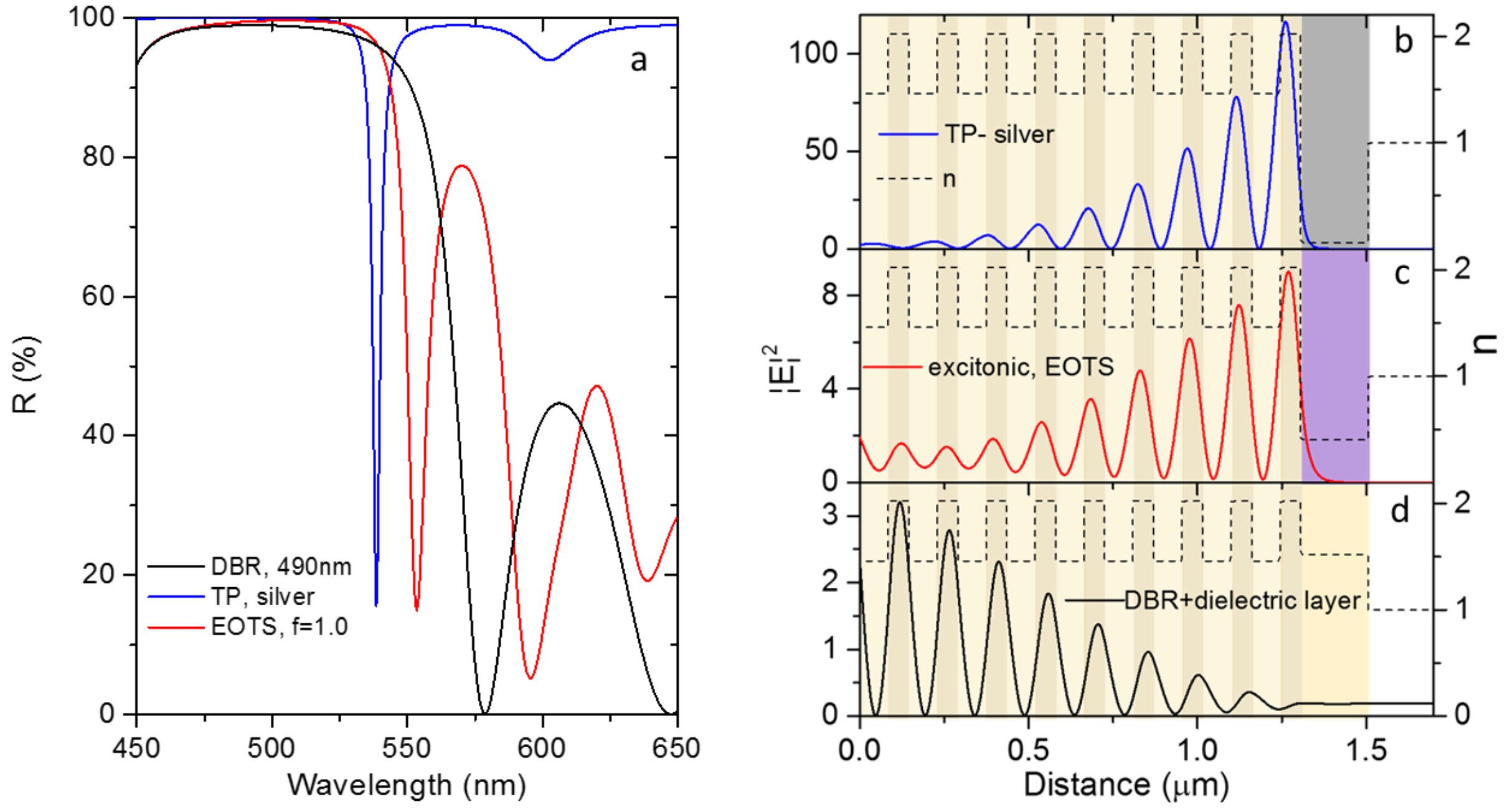

Figure 1. a) Simulated P-polarized (TM) reflection at normal incidence of: a single DBR stack ($\lambda_C$=490 nm, black line), an EOTS structure (red line) with an excitonic thin film with an oscillator strength value of 1 and a TP structure (blue line) with silver as metal top layer. Electric field intensity profile in the three structures at normal incidence for: b) TP mode ($\lambda$=538 nm, blue line), c) EOTS mode ($\lambda$=553 nm, red line), d) stopband region ($\lambda$=553 nm, black line). Black dotted lines represent the real part of the refractive index profile. Different colour regions represent the layers with different optical properties. DBR is represented by cream ($SiO_2$,n=1.46) and light brown ($SiN_3$, n=2.02). Top layer is shown as grey for silver (n=0.06), lilac for excitonic material (n=0.40) and light yellow as no doped PVA (n=1.52). White colour represents air (n=1).

Figure 2

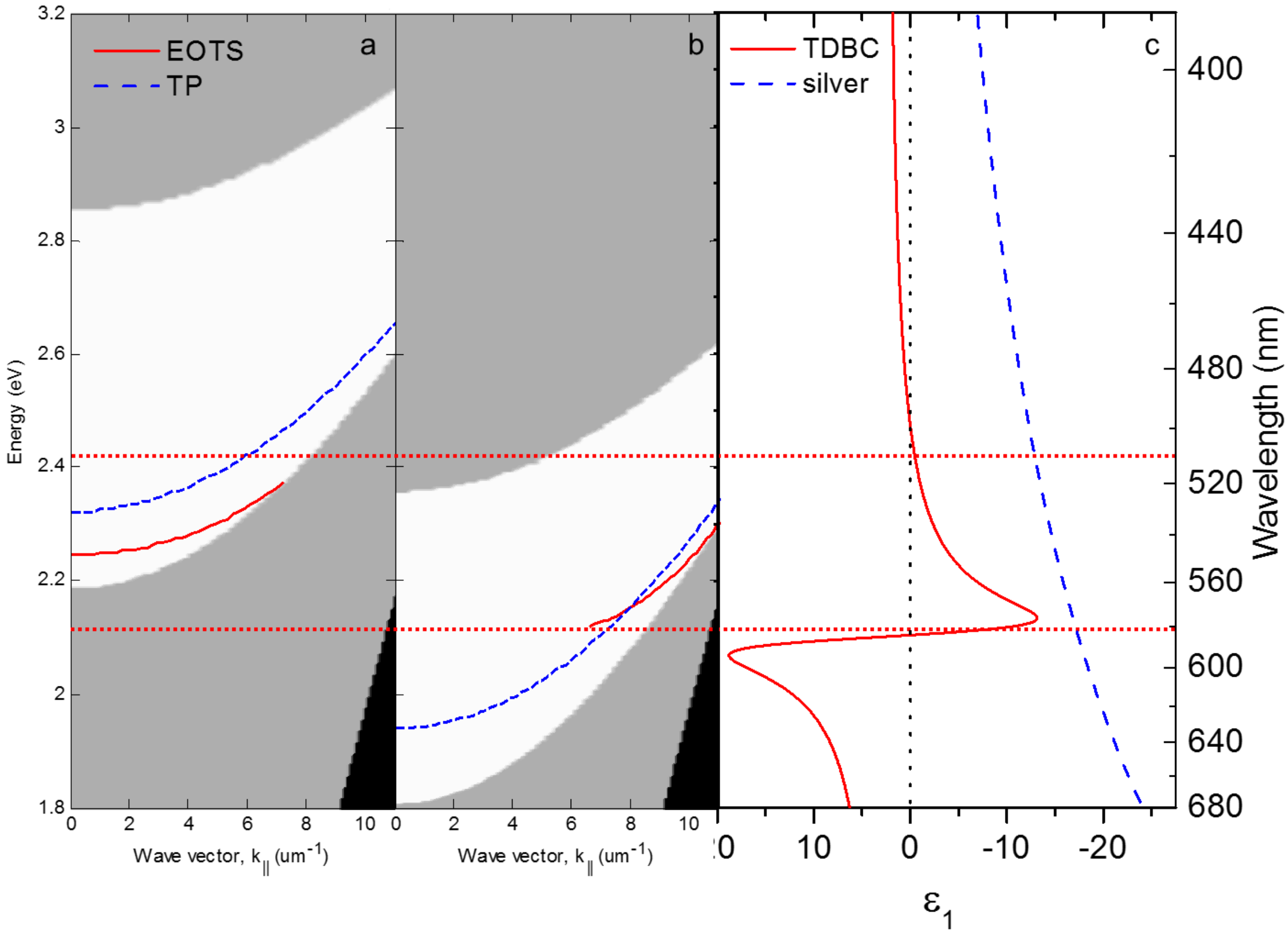

Figure 2. a-b) DBR stop band representation for P-polarized light (TM) for two different DBR's with central wavelengths of 490 nm in 2.a and 590 nm in 2.b. White areas represent the stopband region where the reflectance is higher than 70%. The black regions correspond to areas below the light-line which are not directly accessible by incident light. The grey areas represent all the regions outside the stopband where the reflectance is lower than 70%. Continuous red lines represent the dispersion curves for the EOTS modes supported by the DBR topped with a layer of 200 nm of excitonic material ($f_0$=1). Blue dashed lines represent the TP dispersion for the same DBR topped by 200 nm of silver. c) Values of the real part of the permittivity for a dielectric medium with an strong resonance at 590 nm and an oscillator strength of 1, and for silver from reference [25]. Red dotted lines mark the spectral region where the real part of the permittivity of the excitonic layer reach negative values of the permittivity.

Figure 3

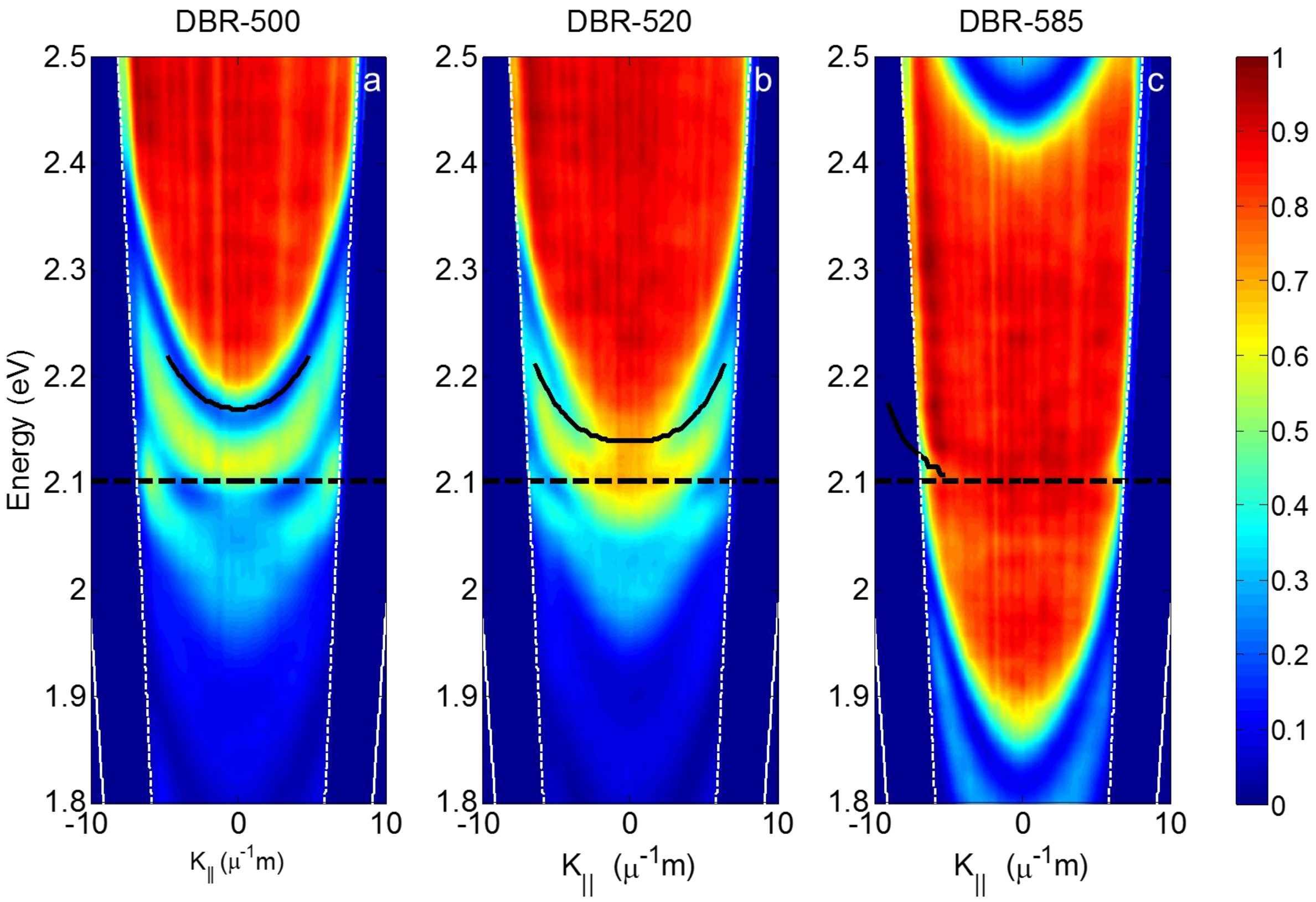

Figure 3. Experimental P-polarized (TM) reflectance maps for: a) EOTS structure composed of a DBR with a central wavelength of 500 nm, b) EOTS structure composed of a DBR with a central wavelength of 520 nm, c) EOTS structure composed of a DBR with a central wavelength of 585 nm. Black line curves corresponds to the calculated dispersion curves for the EOTS TM modes. Black dashed lines indicate the exciton absorption resonance of the J-aggregate molecules at 2.1eV. White dashed lines represent the numerical aperture of the objective used in the reflectance collection and illumination. The continuous white lines correspond to the light cone. In 3.c only the dispersion curve in the negative quadrant is displayed so as to make visualization of the measured features in the positive quadrant of the figure clearer.

Figure 4

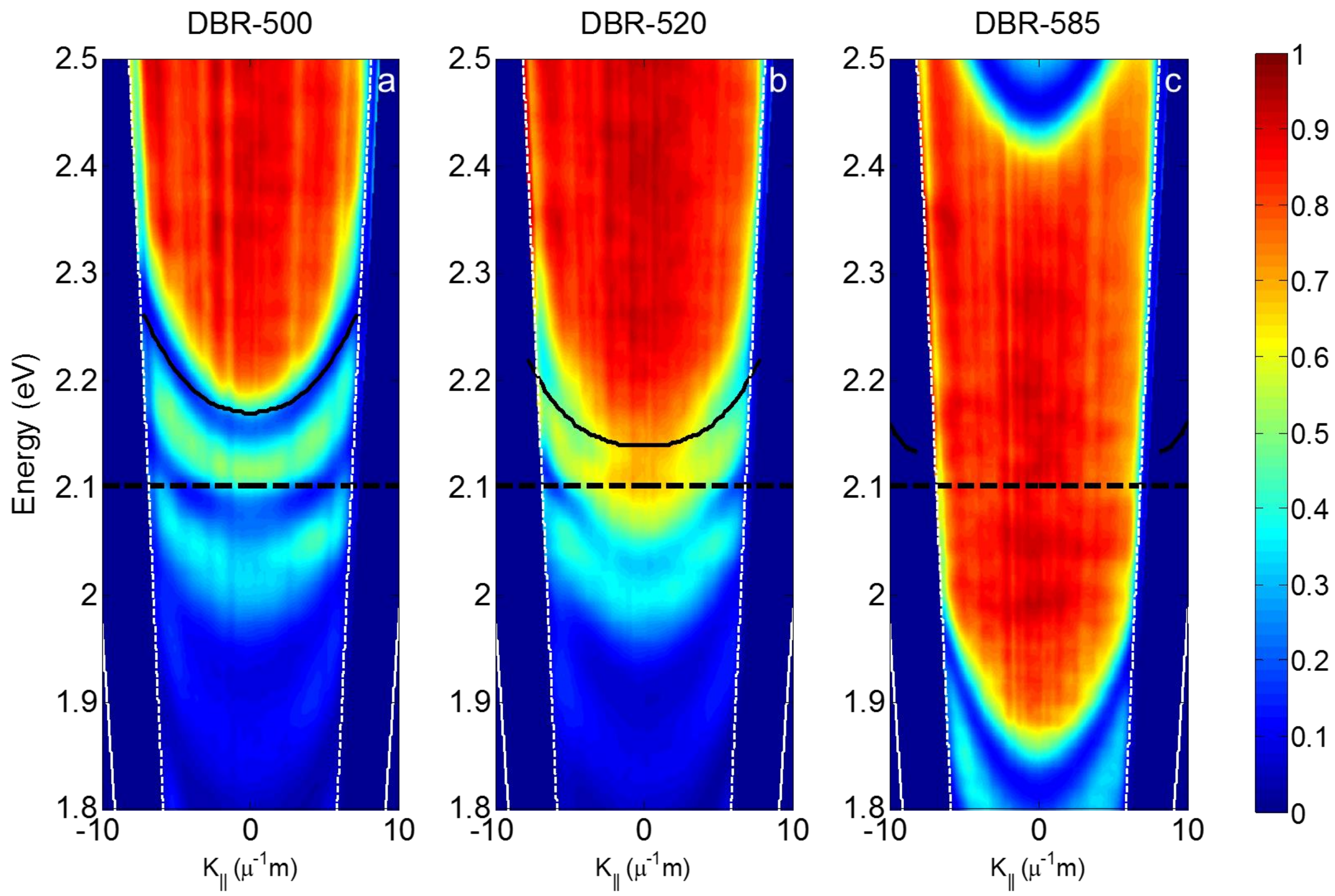

Figure 4. Experimental S-polarized (TE) reflectance maps for: a) EOTS structure composed of a DBR with a central wavelength of 500 nm, b) EOTS structure composed of a DBR with a central wavelength of 520 nm, c) EOTS structure composed of a DBR with a central wavelength of 585 nm. Black line curves corresponds to the calculated dispersion curves for the EOTS TE modes. Black dashed lines indicates the exciton absorption resonance of the J-aggregate molecules at 2.1eV. White dashed lines represent the numerical aperture of the objective use in the reflectance collection and illumination. The continuous white lines correspond to the light cone.

## SUPPLEMENTARY INFORMATION

**Manuscript entitled "Evidence of Excitonic Optical Tamm States using Molecular Materials"**

**Authors:** S. Núñez-Sánchez, M. López-García, M. M. Murshidy, A. G. Abdel-Hady, M. Y. Serry, A. M. Adawi, J.G. Rarity, R. Oulton, W.L. Barnes

1.-High reflectance in heavily doped polymer films as a function of oscillator strength

Observation of a metallic lustre in a thin film is possible when the value of the real part of the permittivity of the material is negative enough. In a material described by a Lorentzian expression the restricted wavelength range where the real part of the permittivity of a material is negative to show a high reflectance response, depends strongly on the oscillator strength. The figure 1 shows the simulated reflection using an iterative recursive Fresnell method for two different thin films with different oscillator strengths. The figure 1.a corresponds to the optical properties used in the modelling part of the paper ($f_0$=1). Figure 1.b corresponds to the equivalent model for our experimental thin excitonic thin films ($f_0$=0.4).

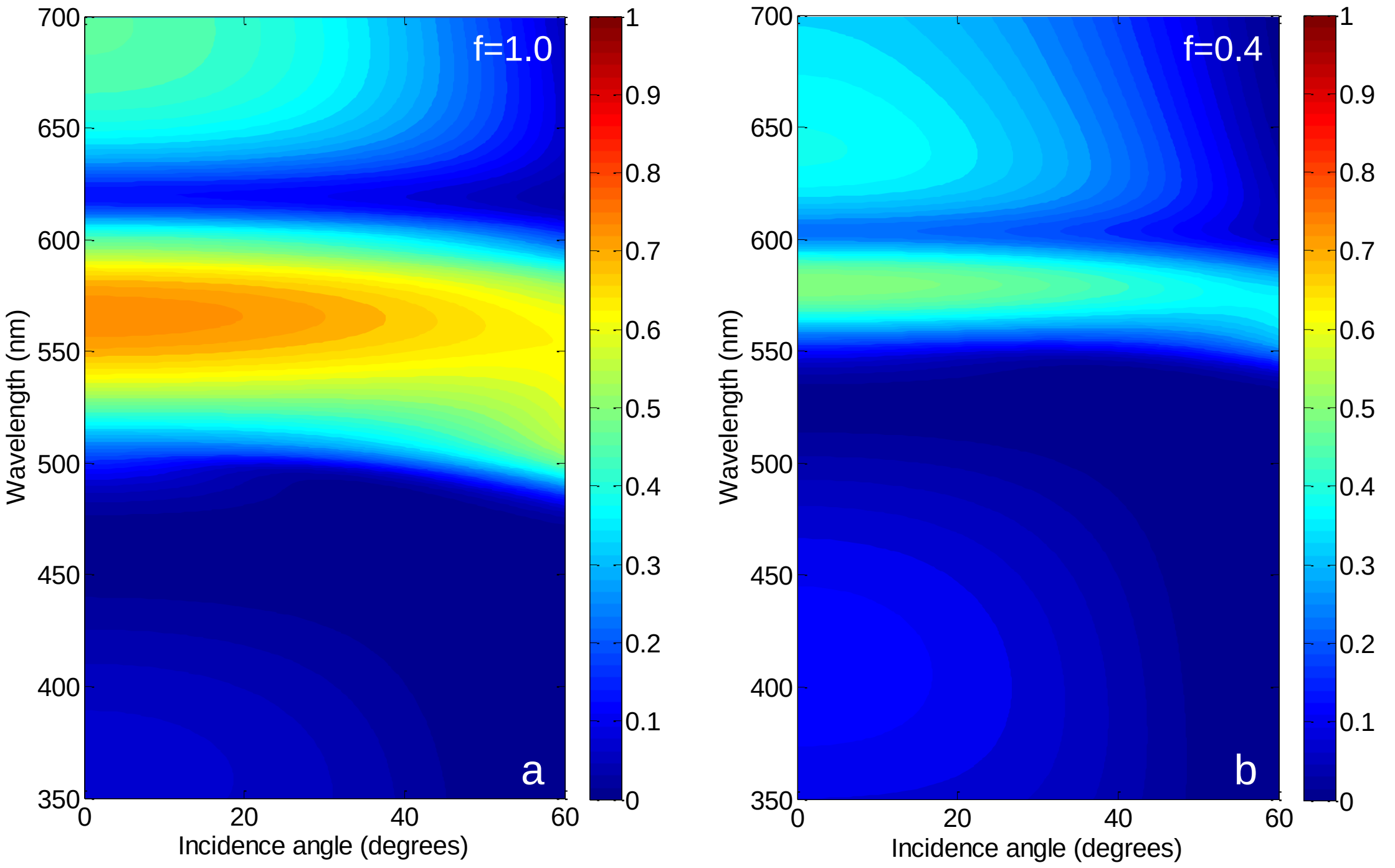

*Figure 1. P-polarized reflectance for a 200 nm thin film with two different oscillator strengths: a) f=1 b) f=0.4.*

2.- Determination and analysis of the dispersion curves of Exciton Optical Tamm States (EOTS)

2.1- Determination of dispersion curves by reflectance simulations

We have studied different methods in order to obtain the dispersion curves of the EOTS and to determine the cut-off conditions. The first method used is based on a qualitative observation on the comparison of the optical response of a bare DBR and the EOTS supporting structure (DBR covered with the J-aggregate layer). The Figure 2 shows the simulated reflectance at different angles obtained by recursive Fresnel formulation for a bare DBR and an EOTS-structure. The DBR is formed by 9 pairs of SiO2/SiN3 layers with a central wavelength of 520 nm. As it is well known, the optical response of a bare DBR shows a reflectance values close to the 100% in the stop band wavelength range with Fabry-Perot oscillations in the laterals. However, in case of EOTS, a dip in reflectance is observed at those wavelengths for which excitation of the OTS is possible. This condition can be only fulfilled at wavelengths lower than the low energy band edge of the bare DBR.  Therefore, we may be able determine the mode cut-off wavelengths simply following the position of the dip in the reflectance inside the stop band of the bare DBR.

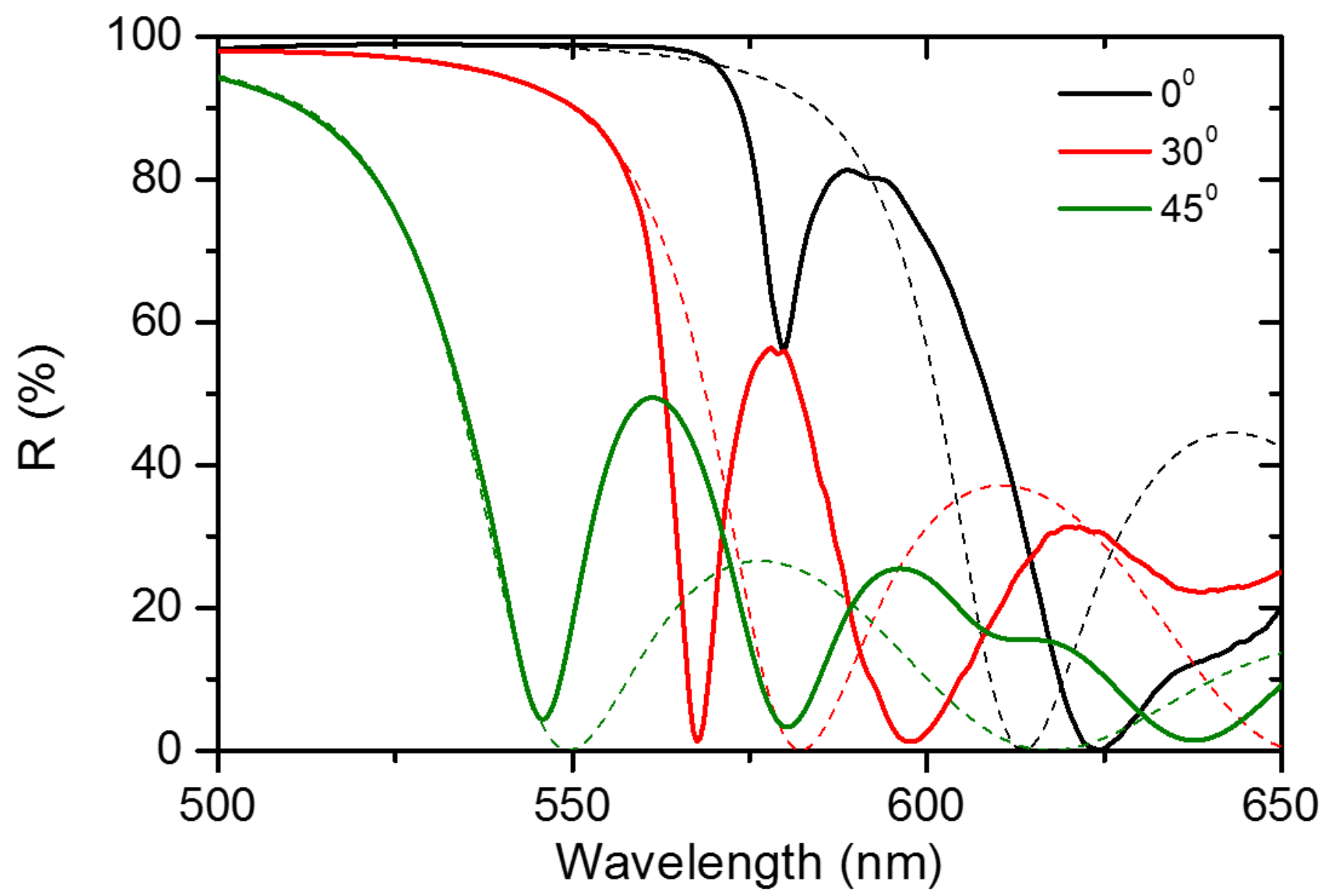

*Figure 2. P-polarized reflectance of a DBR and an EOTS-structure using a DBR formed by 9 pairs of SiO2/SiN3 layers of with a central wavelength of 520 nm. The optical properties of the 200 nm excitonic layer correspond to the real properties or our excitonic material. Labels indicate the incidence angle.*

We have established a numerical method to determine the dispersion curves by reflectance simulations. We have defined a new function that is the subtraction of the reflectance of the bare DBR from the reflectance  of the EOTS-structure  (Subtraction($\lambda$,angle)=$R_{DBR}$-$R_{EOTS}$). The Figure 3 shows the subtraction function as a function of the wavelength and the reflectance for the bare DBR and the EOTS structure at 20 degrees. When the mode condition is fulfilled, the subtraction function will have a local maximum. In our method we determine the wavelength associated to the local maxima of the substraction function at different incident angles, only taking into account wavelengths within DBR stop band.

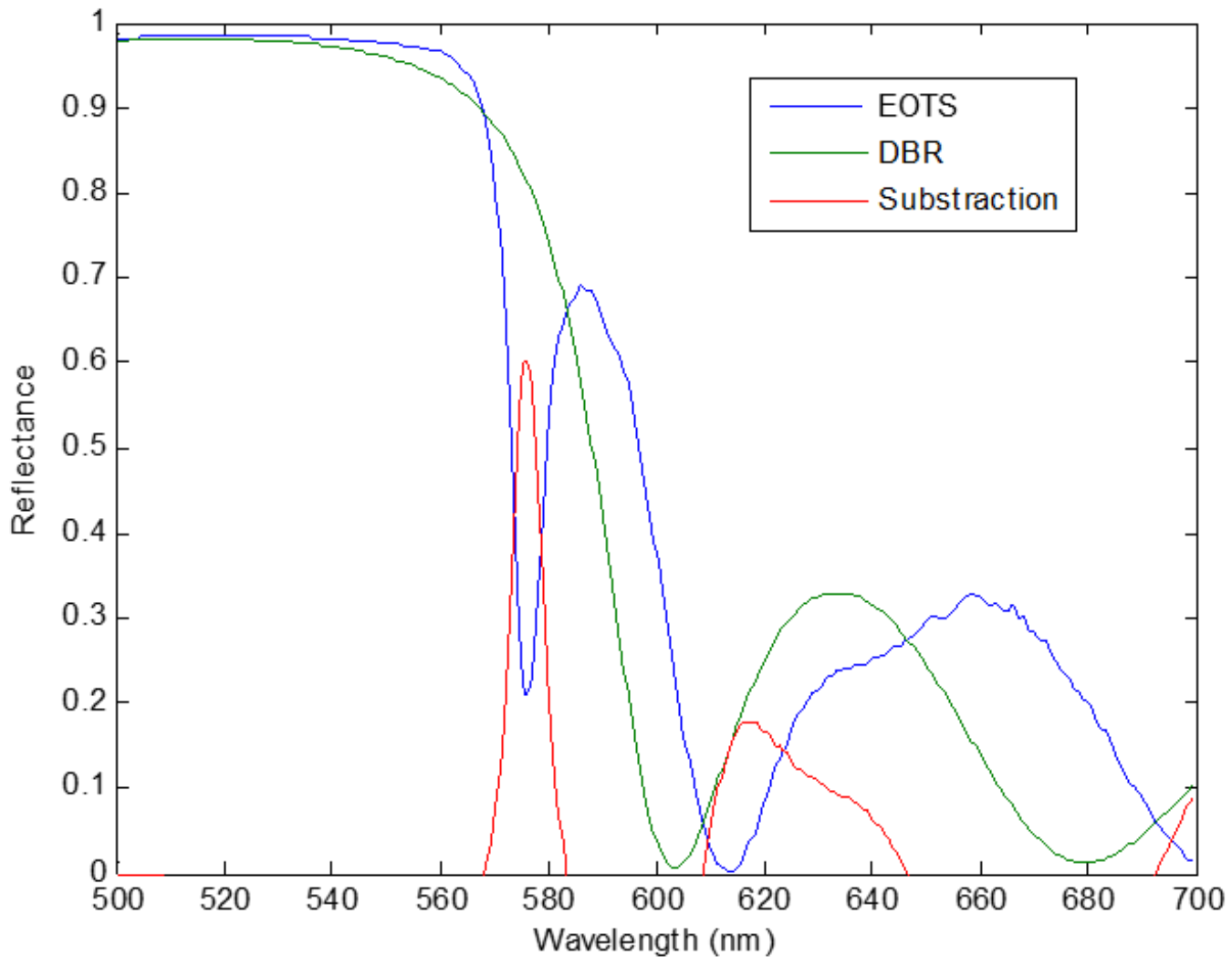

*Figure 3. P-polarized reflectance at 20 degrees for a DBR with a central wavelength of 520 nm and an EOTS structure with a 200nm of our TDBC material on top the DBR. The value of the subtraction function has been also included.*

The contour plots of the figure 4 show the reflectance values obtained for a DBR with a central wavelength of 520 nm, the associated EOTS structure and the substation function as a function of the wavelength and incident angle.  We have superposed a red curve that corresponds to the dispersion curved obtained following the local maxima of the substation. In this case, the cut-off wavelengths at shorter wavelengths is mainly established by the band edge of the DBR.

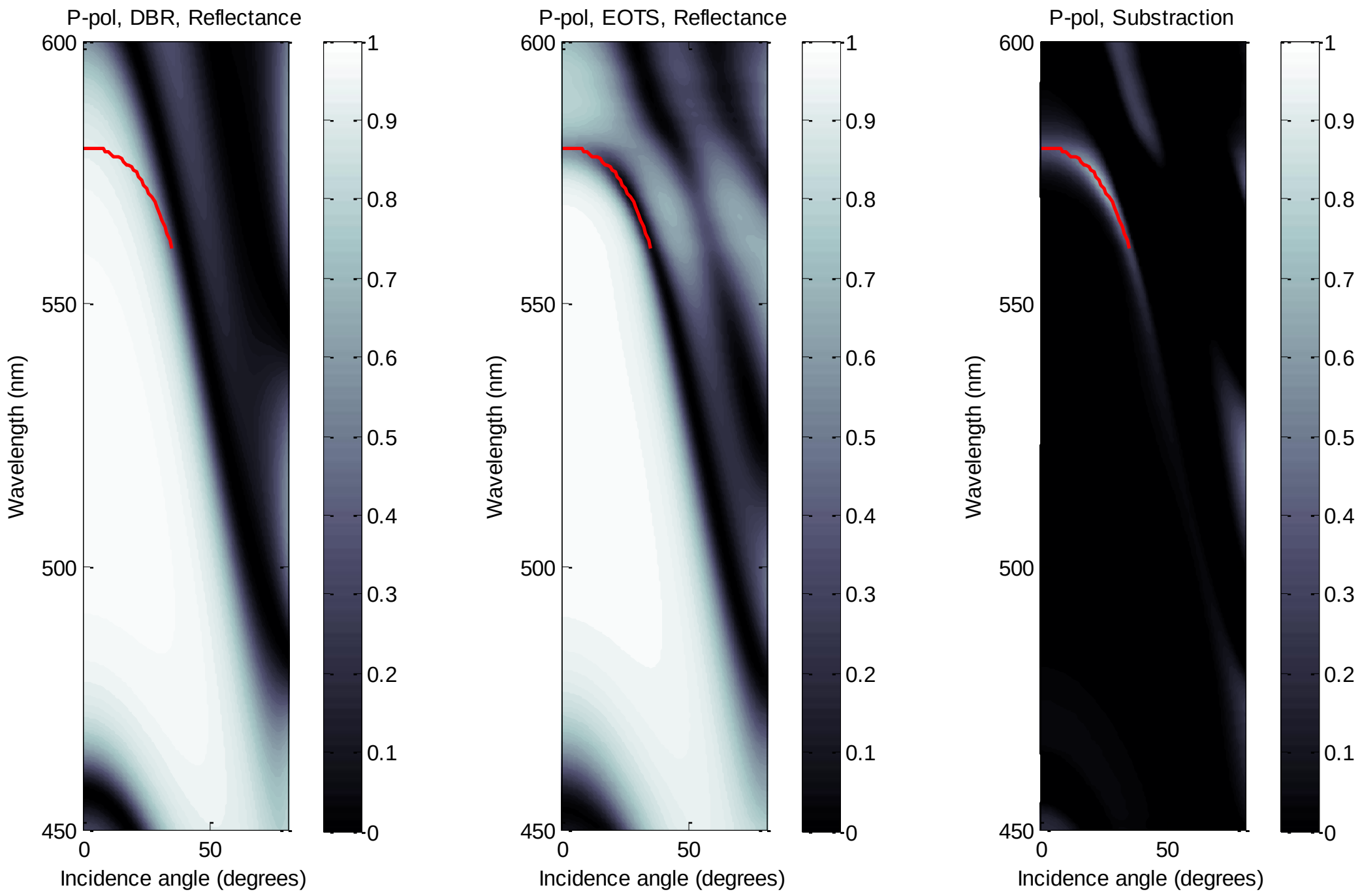

*Figure 4. Contour plot of the P polarized reflectance of a) DBR formed by 9 layers of SiO2/SiN3, b) EOTS structure with the same DBR. c) Values of the substraction function for the same structures. The red line corresponds to the estimated dispersion curves.*

2.2. Determination of dispersion curves by mode condition equation

From Kaliteevski et al [1] we know that the condition in the interface DBR-thin film for an eigenmode associated to a Tamm Optical State is: $r_{left} r_{right} = 1$. Where $r_{left}$ and $r_{right}$ are the Fresnell amplitude coefficients at the interface DBR-thin film. Therefore we could estimate the dispersion curves from the two Fresnel amplitude coefficients at different angles and analysing at which wavelengths the product of the two is close to one.

This equation has been established for an ideal situation, that is, two perfect mirrors with reflectance of 100%. However, in our case the DBR is formed by only 9 layers and the thin film shows a high reflectance, but never close to the total reflection condition (usually around 60% as maximum depending of the angle, see Figure 1). Therefore, we decided to estimate the dispersion curve as the position of the local maximum of the product=$r_{left}$*$r_{right}$ (see Figure 5).

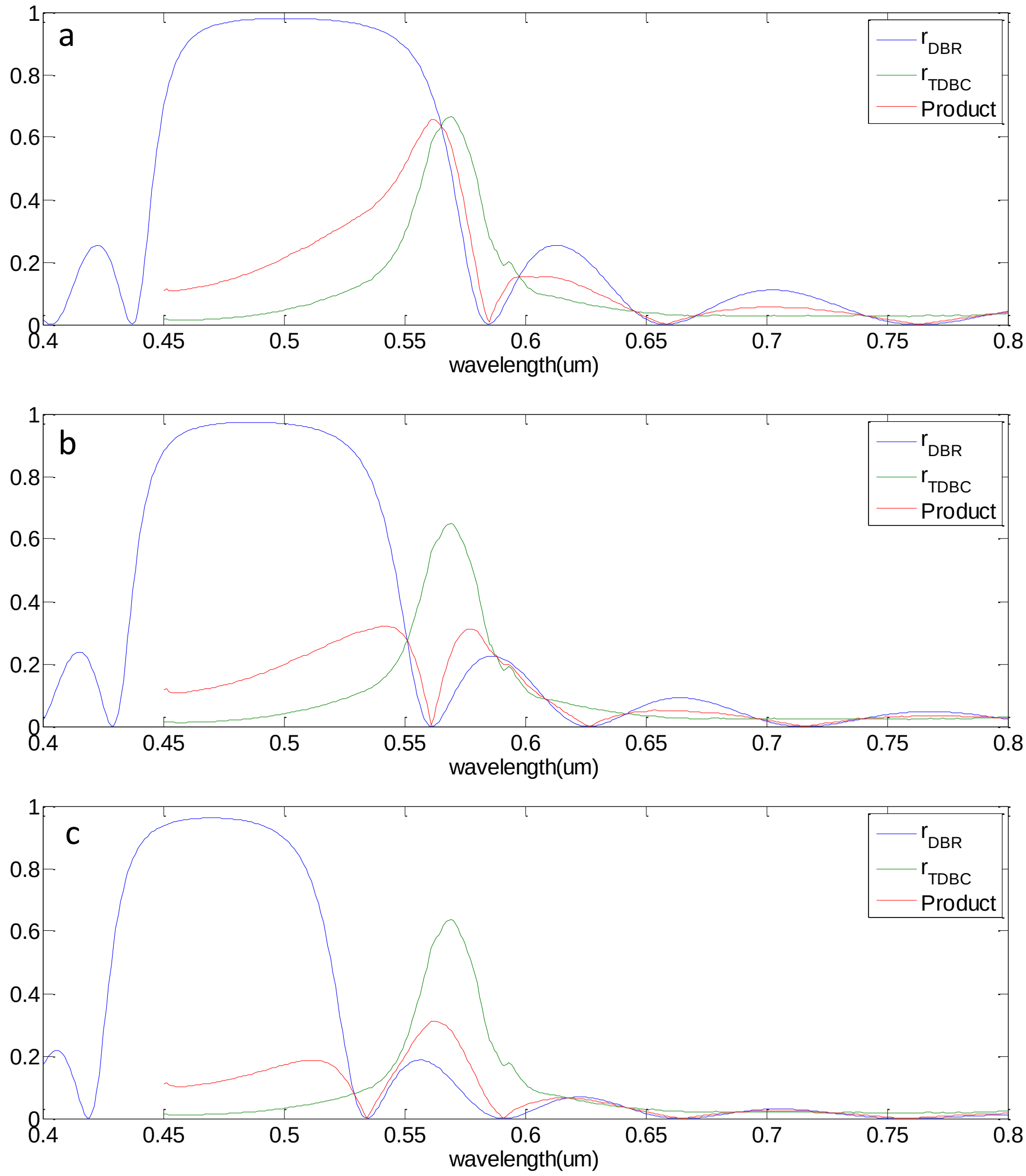

*Figure 5. Amplitude of the Fresnel coefficients for a bare DBR with a central wavelength of 500 nm ($r_{right}$), a EOTS structure formed by the same DBR and 200 nm of a excitonic thin film ($r_{left}$) and the product of the two coefficients at different incidence angles. a) normal incidence, b) 26 degrees, c) 40 degrees.*

The product function of the amplitude of the two Fresnel coefficients shows a minimum around the band edge position of the DBR. Specifically the product function shows two local maxima for larger and shorter wavelengths than the DBR band edge. We observe than when the mode condition is fulfilled (as at normal incidence, Figure 5a) the absolute maximum of the function product is located for wavelengths bellow the band edge. However, when the EOTS is not excitable, as at 40 degrees (see Figure 4c), the absolute maximum of the function product is located at larger wavelengths than the band edge. There is a transition region where the two local maximums are pretty similar, where we expect a hybrid mode (as at 26 degrees, Figure 5b). Therefore we could estimate at different angles (or wave vectors) the mode condition as the wavelength values where the local maximum of the product function is below the DBR band edge.

### 2.3. - Comparison of the dispersion curves obtained for the two methods with the experimental data

If we compare the dispersion curves obtained for the same EOTS structure but using the two different methods, we could observe that the two curves are similar but not exactly the same (Figure 10). The main difference is that the dispersion curve obtained from the product establish the mode condition at higher energies than the one by reflection. This difference is because product method searches the shorter wavelength where the mode condition is fulfilled, therefore it will follow the edge of the feature of the dip in reflectance. However the mode condition estimated by reflection follows the minimum in reflectance. Independently on this difference cut-off conditions are equivalent. In all the figures of the paper we have decided only show the dispersion curves obtained from the reflection simulations.

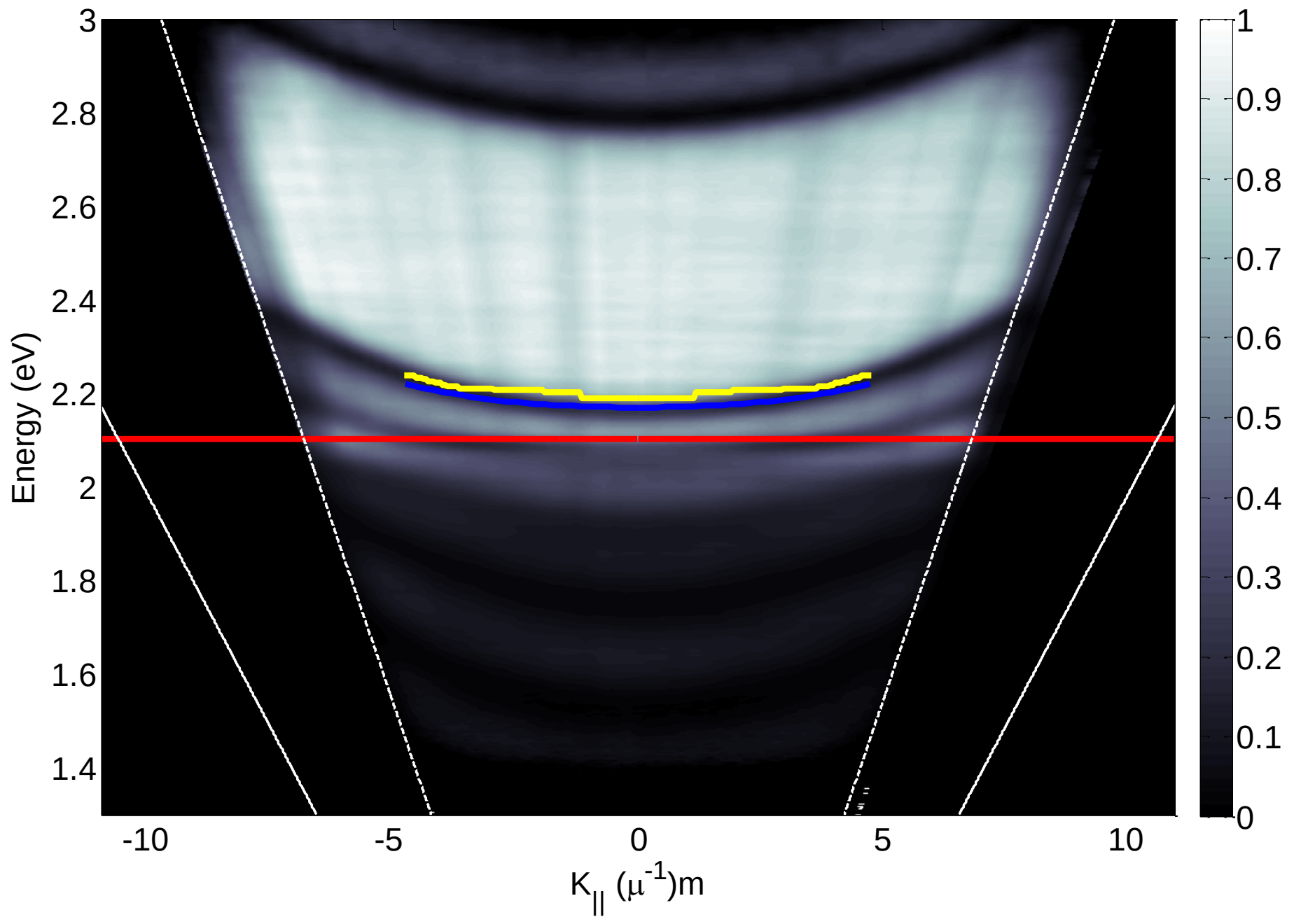

*Figure 6. Experimental reflectance of the EOTS structure with a DBR band centre at 500 nm. The dispersion curve obtained from substraction function of the EOTS simulations and the bare DBR is shown in blue color. The yellow line is the dispersion curve estimated by the location of the maximum of the product of the Fresnel coefficients.*

3- Verification of the obtained wavelength cut-off values. Analysis of field profiles.

In order to verify that our analysis and estimation of the cut-off wavelengths of the EOTS was correct we have plotted the intensity field profiles for the EOTS modes and compare with the intensity field profiles of the band edge mode of the bare DBR at the same angle. The figure 7 shows the reflectance values for a DBR with a central band at 500 nm and its equivalent EOTS structure. The estimated EOTS dispersion curve shows a cut-off wavelength of 557 nm at 26 degrees, determined when the substraction function cross the bare DBR edge. If we analysed the features on the reflectance simulated for the EOTS structure in figure 7.b it is not evident cut off wavelength position. From only reflection simulations of the EOTS (Figure 1.b) it is not clear when mode condition is dismissed and the minimum observed corresponds to the DBR stop band edge.

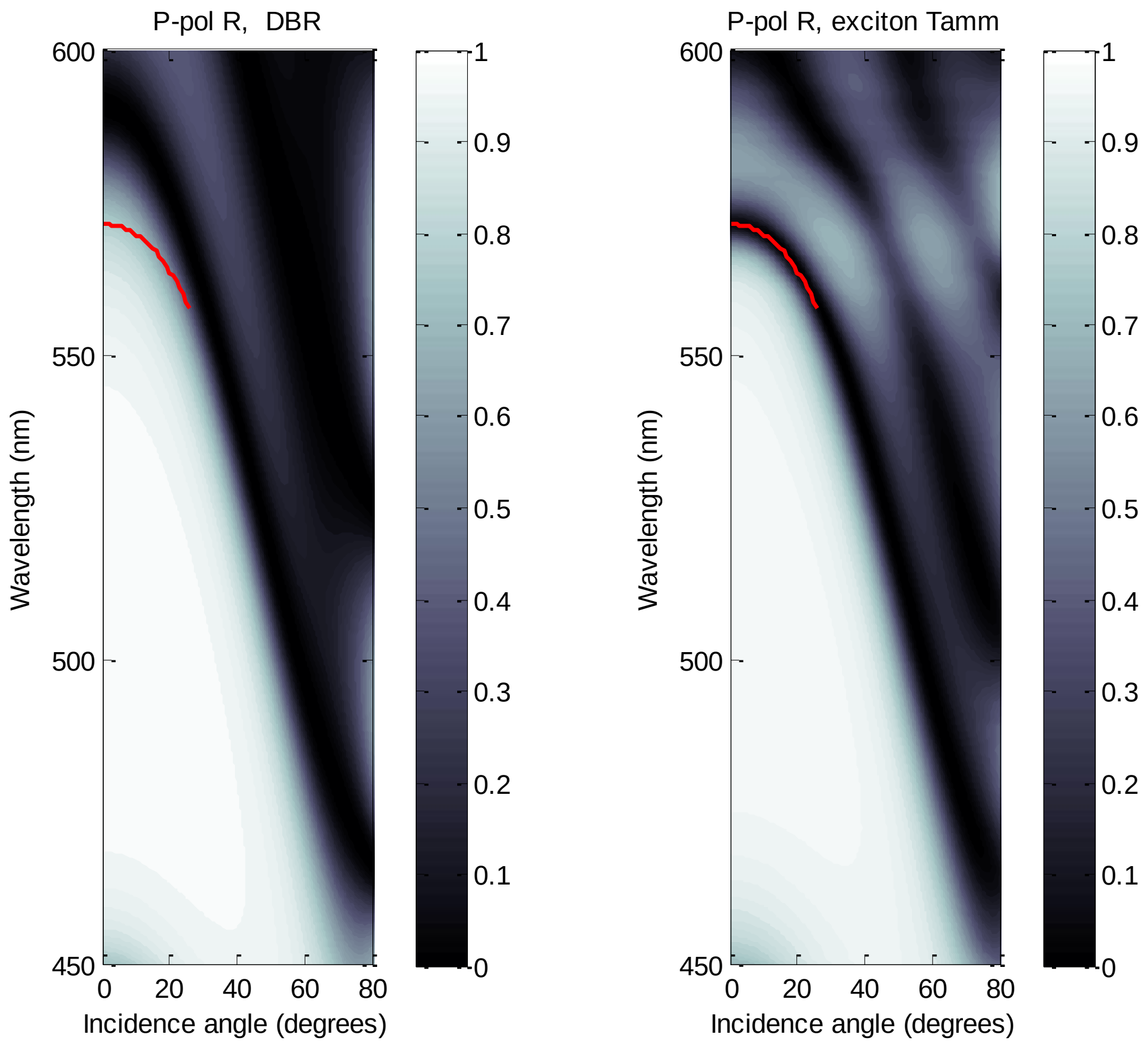

*Figure 7. Contour plot of the P polarized reflectance of a) DBR with a central wavelength of 500 nm, b) EOTS structure with the same DBR and a 200 nm of our excitonic reflector. The red lines correspond to the estimated dispersion curves.*

The electric intensity profiles for the EOTS mode and the band edge mode in the bare DBR at normal incidence are shown in the Figure 8. The intensity profile of the electrical field of an Optical Tamm State along the photonic structure is well establish on the literature. [1,2] The electrical intensity profile decays exponentially in the positive and negative directions from the interface between the DBR and the thin film. The most characteristic features are an oscillating decay within the DBR structure due to the periodic change of the refractive index and a faster decay in the thin film medium. In contrast, DBR edge mode should be a symmetric oscillating field as it can be observed in Figure 8.

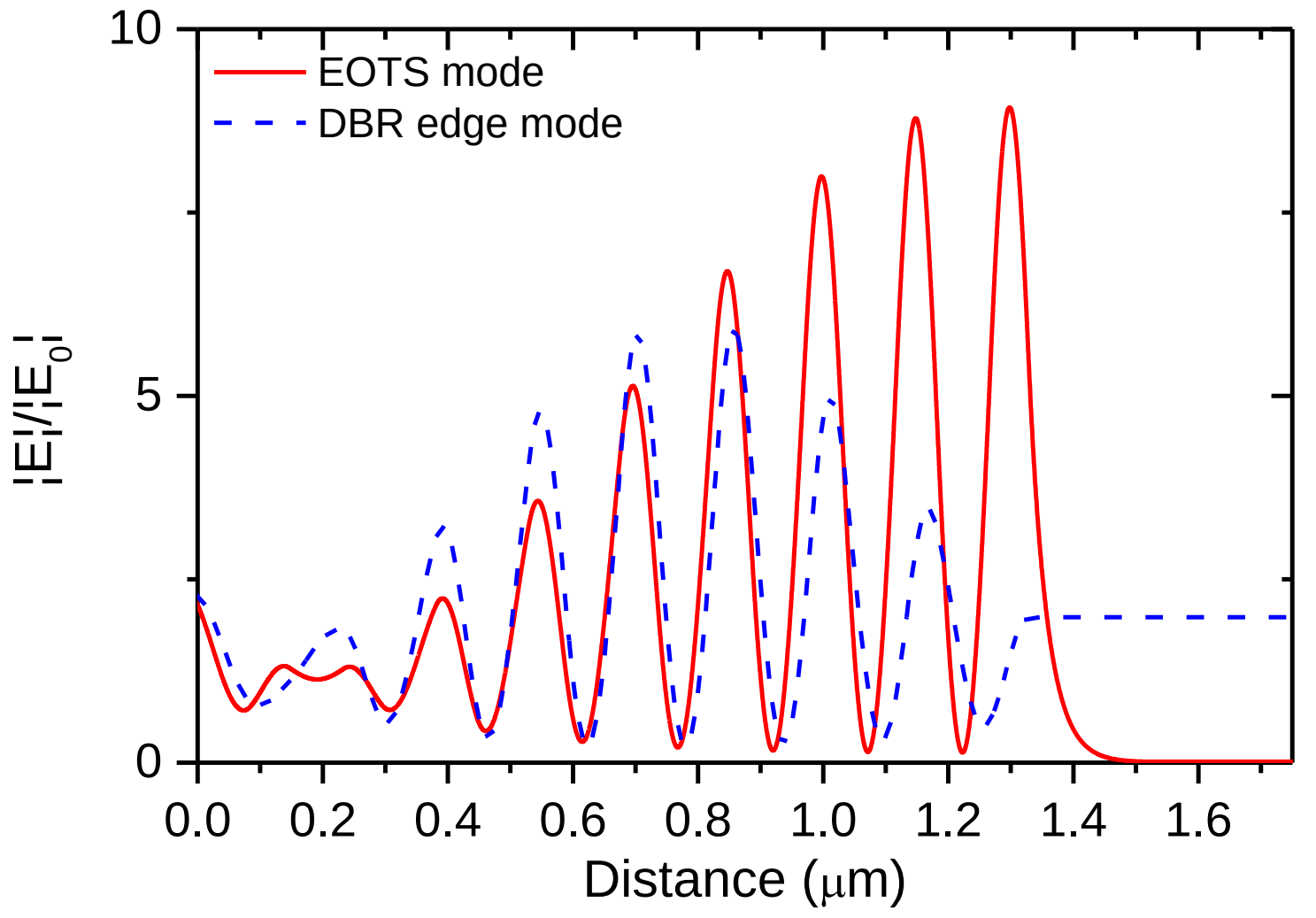

*Figure 8. Electric intensity field profiles at normal incidence for at EOTS (wavelength=571nm) and a DBR edge mode (wavelength=590 nm) at normal incidence.*

For the structure represented on Figure 7, from 0 degrees to 20 degrees the intensity field profiles supported by EOTS structure are similar to the characteristics of a OTS, confirming that the minimum observed corresponds to the EOTS mode condition. However from 20 to 30 degrees the mode transforms into an hybrid mode between a OTS mode and an DBR edge mode (see Figure 9).

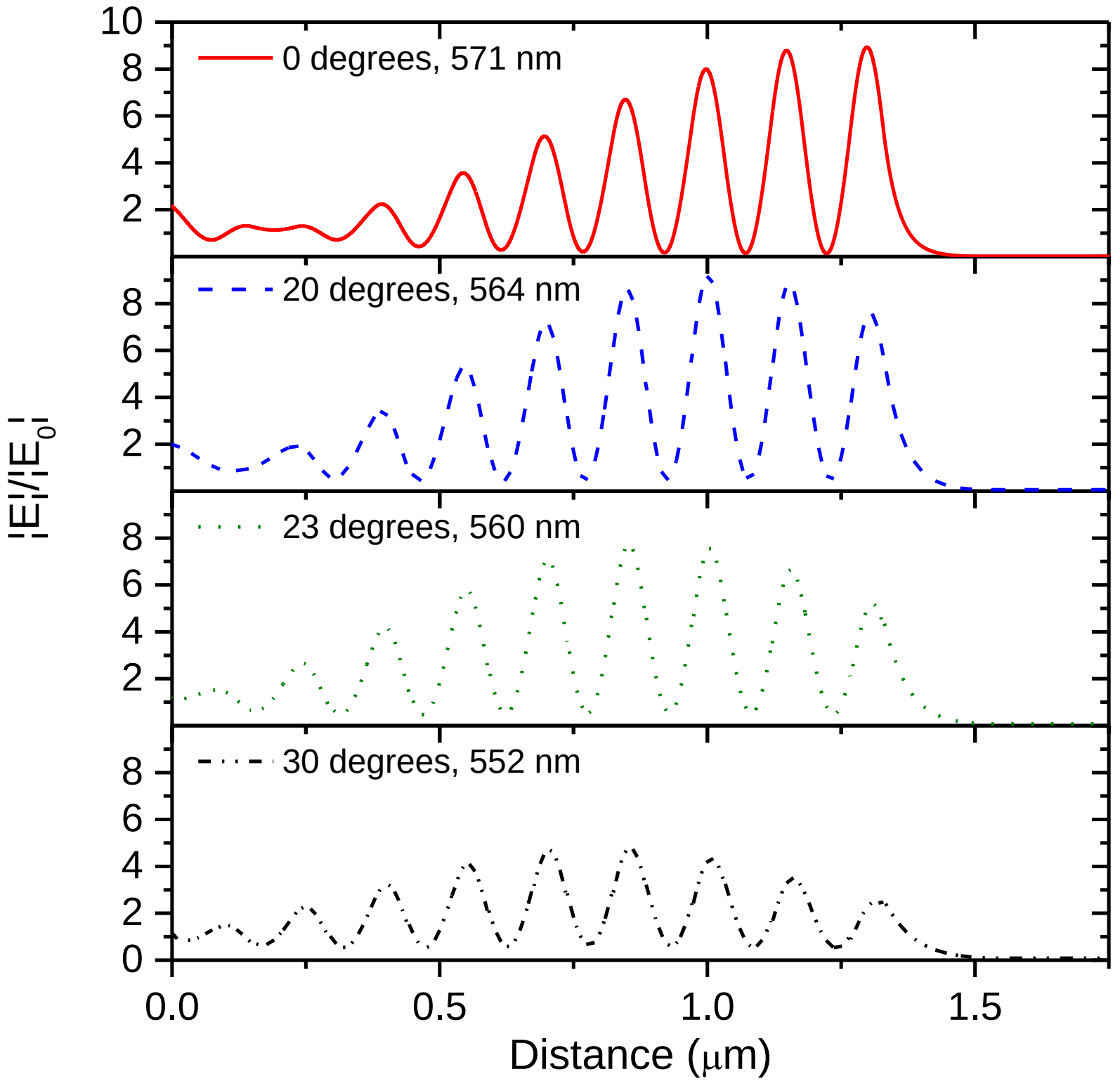

*Figure 9. Electric intensity field profiles for at EOTS structure at different incident angles at mode condition (0, 20, 23 degrees) or DBR edge conditions (30 degrees).*

The cut-off wavelength for this EOTS mode has been estimated around 26 degrees at 557 nm The mode associated to the minimum observed in the reflectance of the EOTS structure for larger angles and shorter wavelengths follows clearly the tendency of an edge mode without any exponential decay envelope, as it is shown in Figure 9 at 30 degrees.

## 4. Splitting between TE and TM modes of EOTS

As we mentioned on the paper, OTS are characterized by the splitting of between TE and TM polarized modes that increases quadratically as a function of the in-plane wave-vector. As OTS, EOTS dispersion curves follows the same tendency, with the difference that in the EOTS, the limited metal-like band causes different cut-off wavelengths between TE and TM modes for the same structure (see Figure 10).

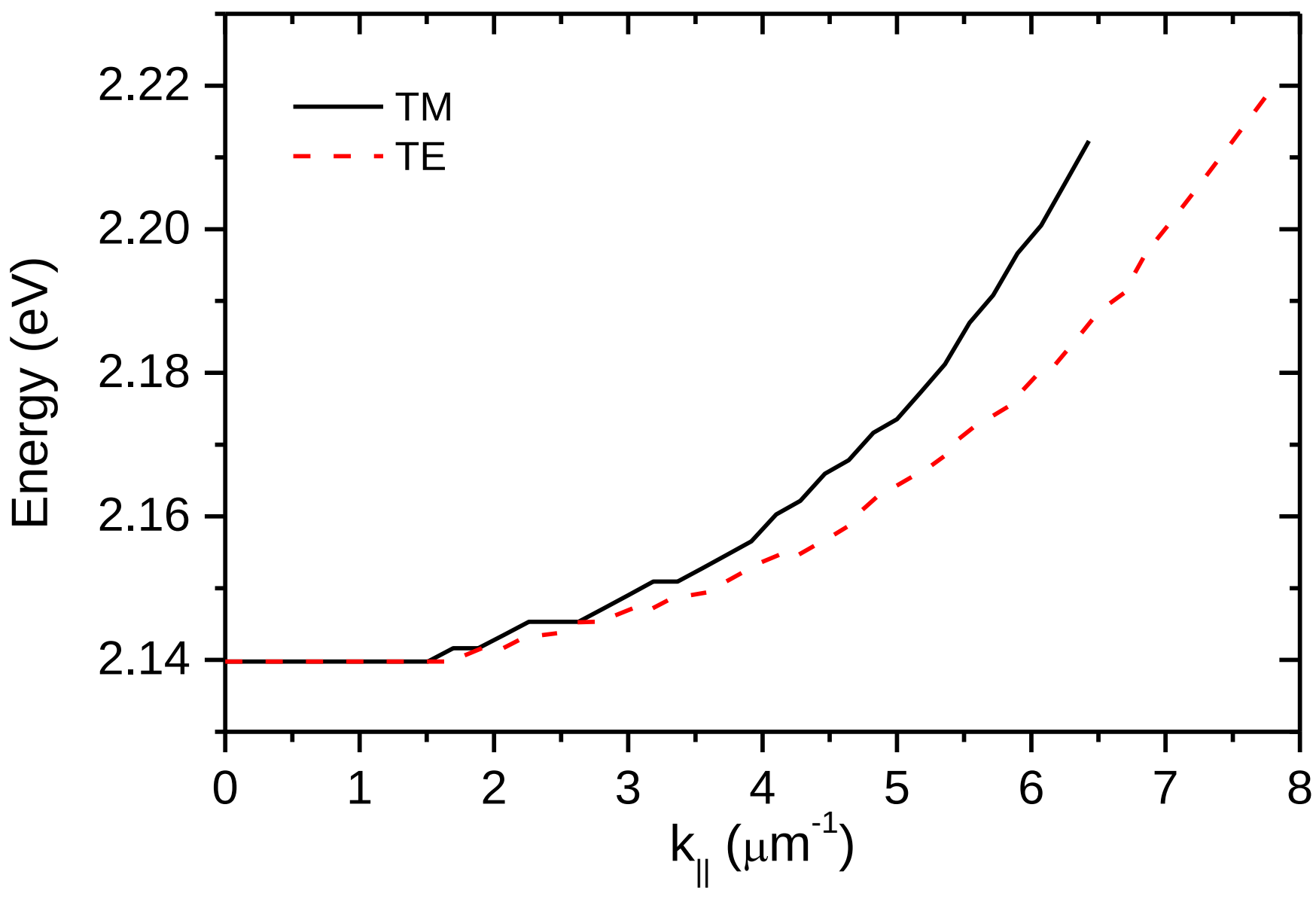

*Figure 10. Dispersion curves of TE and TM modes estimated by reflection simulations for a EOTS structure formed by a DBR with a central wavelength of 520 nm and an excitonic layer of 200 nm.*